\documentclass[10pt,twoside]{article}

\usepackage{fancyhdr}
\usepackage{titlesec}
\usepackage{cite}
\usepackage{ifthen}
\usepackage{graphicx}
\usepackage{float}
\usepackage{amsmath,amssymb}

\titleformat{\section}{\centering\large\bfseries}{\S\arabic{section}}{1em}{}
\newboolean{first}
\setboolean{first}{true}

\begin{document}

\setlength\abovedisplayskip{2pt}
\setlength\abovedisplayshortskip{0pt}
\setlength\belowdisplayskip{2pt}
\setlength\belowdisplayshortskip{0pt}

\title{\bf \Large  Multiple Vectors Propagation of Epidemics in Complex Networks
\author{Dawei Zhao$^{a,b}$   \  \ Lixiang Li$^{a,b}$   \  \ Haipeng Peng$^{a,b}$   \  \ Qun Luo$^{a,b}$\ \ Yixian Yang$^{a,b}$ \\ \small \it $^{a}$Information Security Center, Beijing University of Posts and
Telecommunications, \\
\small \it Beijing 100876, China. \\ \small \it $^{b}$National
Engineering Laboratory for Disaster Backup and
Recovery, \\
\small \it Beijing University of Posts and Telecommunications,
Beijing 100876, China. }\date{}} \maketitle

\footnote{E-mail address: dwzhao@ymail.com (Dawei Zhao);
penghaipeng@bupt.edu.cn (Haipeng Peng).}

\begin{center}
\begin{minipage}{135mm}
{\bf \small Abstract}.\hskip 2mm {\small This letter investigates
the epidemic spreading in two-vectors propagation network (TPN). We
propose detailed theoretical analysis that allows us to accurately
calculate the epidemic threshold and outbreak size. It is found that
the epidemics can spread across the TPN even if two
sub-single-vector propagation networks (SPNs) of TPN are well below
their respective epidemic thresholds. Strong positive degree-degree
correlation of nodes in TPN could lead to a much lower epidemic
threshold and a relatively smaller outbreak size. However, the
average similarity between the neighbors from different SPNs of
nodes has no effect on the epidemic threshold and outbreak size.}
\end{minipage}\end{center}
\begin{center}
\begin{minipage}{135mm}
\end{minipage}
\end{center}

\section{Introduction}
\label{}

In recent years, various types of epidemics have occurred frequently
and spread around the world, causing not only a great economic loss,
but also widespread public alarm. For example, the intense outbreak
of SARS caused 8,098 reported cases and 774 deaths. Within weeks,
SARS spread from Hong Kong to infect individuals in 37 countries in
early 2003 [1]. An outbreak of mobile viruses occurred in China in
2010. The `Zombie' virus attacked more than 1 million smart phones,
and created a loss of \$300,000 per day [2]. And we have also
witnessed how social networks being used for citizens to share
information and gain international support in the Arab Spring [3].
In view of these situations, it is thus urgent and essential to have
a better understanding of epidemic process, and to design effective
and efficient mechanisms for the restraint or acceleration of
epidemic spreading.

Valid epidemic spreading models can be used to estimate the scale of
a epidemic outbreak before it actually occurs in reality and
evaluate new and/or improved countermeasures for the restraint or
acceleration of epidemic spreading. In the last decade, there have
been extensive studies on the modeling of epidemic dynamics [4-10]
and various protection strategies have been proposed and evaluated
[11-17]. However, these existing researches have been dominantly
focusing on the cases that epidemics spread through only one vector.
While in reality, many epidemics can spread through multiple vectors
simultaneously. For example, it has been well recognized that AIDS
can be transmitted via vectors such as sexual activity, blood and
breast milk; rumor or information can be spread among groups through
verbal communication and social networks; malwares can move to
computers by P2P file share, email, random-scanning or instant
messenger [18]; some mobile malwares can even attack smart phones
through both short messaging service and bluetooth at the same time
[19]. In this letter, epidemic spreading via only one vector and
that through various vectors are called single-vector propagation
and multiple-vectors propagation respectively. Obviously, the range
and intensity of the multiple-vectors propagation will be greater
than the traditional single-vector propagation. Besides, different
propagation vectors can form different propagation networks and
these propagation networks may have different topological
characteristics and epidemic dynamics. Based on the above analysis,
the study of multiple-vectors propagation of epidemics is definitely
a very meaningful and necessary thing.

To the best of our knowledge, a theory describing the
multiple-vectors propagation of epidemics has not been fully
developed yet. In this letter, we propose and evaluate two-vectors
propagation of epidemics in a two-vectors propagation network (TPN)
following the typical Susceptible-Infected-Removed (SIR) model
[6,7]. We map the SIR model into bond percolation [7] and develop
equations which allow accurate calculations of epidemic threshold
[6] and outbreak size [6] of the TPN. It is obviously found the
epidemic can spread across the TPN even when the two
sub-single-vector propagation networks (SPNs) of TPN are well below
their respective epidemic thresholds. We also introduce two
quantities for measuring the level of inter-similarity between the
two SPNs. One is ASN, which measures the average similarity between
the neighbors from different SPNs of nodes in the TPN. We find that
epidemic threshold and outbreak size are not significantly affected
by the ASN. The second quantity is DDC which describes the
degree-degree correlation of nodes in different SPNs. Positive
values of DDC indicate that high degree nodes in one SPN are also
high degree nodes in the other SPN and vise versa. It is found that
strong positive DDC leads to a clearly lower epidemic threshold and
a relatively smaller outbreak size, independent of the topological
characteristics of the two SPNs.

\section{Models and analysis}
\label{}
\subsection{Network model}
In this section, we propose a two-vectors propagation network (TPN)
which is superposed by two sub-single-vector propagation networks
(SPNs), but it is easily to extend the model to an arbitrary number
of SPNs with any size.

\begin{figure*}[h]
\centering{\includegraphics[scale=0.45,trim=0 0 0 0]{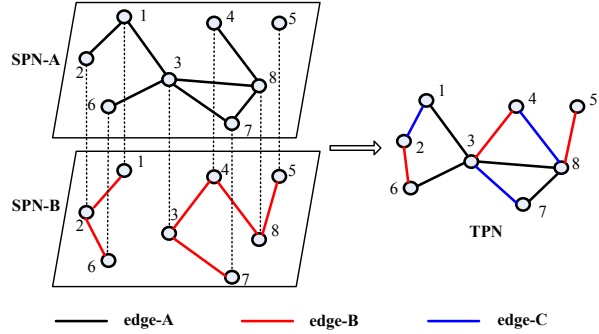}}
\caption{TPN is a superposition network of SPN-A and SPN-B.}
\end{figure*}

As shown in Fig.1, two different spread vectors, vector-A and
vector-B form two SPNs, SPN-A and SPN-B respectively. These two SPNs
have same nodes, but different topologies. When an epidemic can
spread through these two vectors at the same time, the real
propagation network of this epidemic are the TPN (right-hand chart)
which is superposed by SPN-A and SPN-B. Thus, a node in TPN may be
infected by one vector even though it cannot be infected via the
other one, or it may be infected by these two vectors at the same
time. Each node in the TPN has up to three types of edges where the
edge-A belongs to SPN-A, the edge-B belongs to SPN-B,  and the
edge-C belongs to both SPN-A and SPN-B. Vector degree
$k_M\equiv(k_A- k_C, k_B- k_C, k_C)$ is used to characterize the
node of TPN, where $k_A- k_C$, $k_B- k_C$ and $ k_C$ represent the
numbers of edge-A, edge-B and edge-C, respectively. The numerical
value of vector degree of node in TPN is defined by $|k_M|= k_A+k_B-
k_C $. For instance, the vector degree of node 3 of Fig.1 is
$k_M\equiv(4-1, 2-1, 1)$.

Actually, in the vector degree of node in TPN, $k_C$ evaluates how
many neighbors of a node in SPN-A are also its neighbors in SPN-B
which can affect the topology of TPN. We develop a measure,
$\alpha$, called ASN to assess the average similarity between the
neighbors from different SPNs of nodes in the TPN and it is defined
as
\begin{equation}\alpha=\frac{\sum\limits_{i}{k_C}(i)}
{\sum\limits_{i}{|k_M}(i)|},\end{equation} where ${k_C}(i)$ and
$|k_M(i)|$ are the values of $k_C$ and $|k_M|$ of node $i$ in TPN,
respectively. For increasing values of $\alpha$ more of the
neighbors of a node in SPN-A are also its neighbors in SPN-B and
these two SPNs become more similar. For $\alpha=1$, these two SPNs
must be identical.

For a node of TPN, it may be a high degree node in SPN-A and a low
degree one in SPN-B, or a high degree node in SPN-A and also a high
degree one in SPN-B. The influence of different combinations of
nodes degrees in the two SPNs for the characteristics of two-vectors
propagation of epidemics is one of our main research problems.
Analogously to the degree correlation in a single network [20,21]
and the network assortativity in interconnected networks [10], we
define the degree-degree correlation (DDC) of nodes in two SPNs as
follows
\begin{equation}\beta =\frac{\sum\limits_{k_A}\sum\limits_{k_B}(k_A k_B(p(k_A, k_B)-
(\sum\limits_{k_A}p(k_A, k_B))( \sum\limits_{k_B}p(k_A,
k_B))))}{\sum\limits_{k_B}k_B^2 \sum\limits_{k_A} p(k_A,
k_B)-(\sum\limits_{k_B} k_B \sum\limits_{k_A} p(k_A,
k_B))^2},\end{equation} where $p(k_A, k_B)$ denotes the probability
that a randomly chosen node in TPN has degree $k_A$ in SPN-A and
$k_B$ in SPN-B. These two SPNs are said to be disassortative if
$\beta < 0$, assortative if $\beta > 0$, and uncorrelated if $\beta
= 0$.

\subsection{Epidemic spreading model}
The epidemic spreading model adopted here is the
Susceptible-Infected-Removed (SIR) model which is the most basic and
well-studied epidemic spreading model [6,7]. In the SIR model, the
individuals of the network can be divided into three compartments,
including susceptibles (S, those who are prone to be infected),
infectious (I, those who have been infected), and recovered (R,
those who have recovered from the disease). At each time step, a
susceptible node becomes infected with probability $\lambda$ if it
is directly connected to a infected node. The parameter $\lambda$ is
called the spreading rate. Meanwhile, an infected node becomes a
recovered node with probability $\delta$. For the proposed
two-vectors propagation model, we assume that a susceptible node
becomes infected with probabilities $\lambda_A$,$\lambda_B$ and
$\lambda_C$ if it is directly connected to one infected node through
edge-A, edge-B and edge-C, respectively. Obviously,
$\lambda_C=1-(1-\lambda_A)(1-\lambda_B)$. Meanwhile, an infected
node becomes a recovered node with probability $\delta$. Without
loss of generality, we let $\delta=1$.

\subsection{Calculations of epidemic threshold}
Traditionally the percolation process [7] is parametrized by a
probability $\varphi$, which is the probability that a node is
functioning in the network. In technical terms of percolation
theory, one says that the functional nodes are occupied and
$\varphi$ is called the occupation probability. With only slight
modification the general SIR model can be perfectly mapped into the
bond percolation in complex networks where spreading rate
corresponds to the probability that a link is occupied in
percolation [7,10]. We now use the SIR model and the bond
percolation theory to analyze the two-vectors propagation of
epidemics in the defined TPN. For the two-vectors propagation model,
three types of edges are occupied at the probabilities of
$\lambda_A$,$\lambda_B$ and $\lambda_C$ respectively. Let $h_A(x)$
($h_B(x)$, $h_C(x)$) be the generating function [21,22] for the
distribution of the sizes of components which are reached by an edge
with type of edge-A (edge-B, edge-C) and following it to one of its
ends. Later the size of components formed by infected nodes will be
called the outbreak size.
\begin{equation}\begin{split}
\label{cases}h_A(x) = 1-\lambda_A+
x\lambda_A\times\frac{\sum\limits_{k_M,\ k_A-k_C\geq1}
|k_M|p_{k_M}h_A^{k_A-k_C-1}(x)h_B^{k_B-k_C}
(x)h_C^{k_C}(x)}{\sum\limits_{k_M} |k_M|p_{k_M}},\ \
\end{split}\end{equation}
\begin{equation}\begin{split}
\label{cases}h_B(x) = 1-\lambda_B+
x\lambda_B\times\frac{\sum\limits_{k_M,\ k_B-k_C\geq1}
|k_M|p_{k_M}h_A^{k_A-k_C}(x)h_B^{k_B-k_C-1}
(x)h_M^{k_C}(x)}{\sum\limits_{k_M}
|k_M|p_{k_M}},\end{split}\end{equation}
\begin{equation}\begin{split}
\label{cases}h_C(x) = 1-\lambda_C+
x\lambda_C\times\frac{\sum\limits_{k_M,\ k_C\geq1}
|k_M|p_{k_M}h_A^{k_A-k_C}(x)h_B^{k_B-k_C}
(x)h_M^{k_C-1}(x)}{\sum\limits_{k_M}
|k_M|p_{k_M}},\end{split}\end{equation} where $p_{k_M}$ denotes the
probability that a randomly chosen node of TPN has the vector degree
$k_M$.

Generally, an epidemic always starts from a network node, not an
edge, therefore we proceed to analyze the outbreak size distribution
for epidemic sourced\ from\ a randomly\ selected\ node. If we start
at a randomly chosen node in TPN, then we have one such outbreak
size at the end of each edge leaving that node, and hence the
generating function for the outbreak size caused by a network node
is
\begin{equation} \centering{\label{cases}H(x) = x\times\sum\limits_{k_M}
p_{k_M}h_A^{k_A-k_C}(x)h_B^{k_B-k_C} (x)h_C^{k_C}(x).}\end{equation}

Although it is not\ usually\ possible\ to\ find\ a\ closed-form
expression\ for\ the\ complete\ distribution\ of\ outbreak\ size\ in
a network, we can find closed-form expressions for the average
outbreak size of an epidemic in TPN from Eqs.(6). This average
outbreak size can be\ derived\ by\ taking\ derivates\ of Eqs.(6) at
$x = 1$, we have
\begin{equation}\begin{split}
 \label{cases}<s> =H'(1)=
1+\sum\limits_{k_M} p_{k_M}(k_A-k_C)h'_A(1)+\ \ \ \ \ \ \ \ \ \ \ \
\ \ \ \ \ \ \ \ \ \ \ \ \ \ \ \ \ \\
\sum\limits_{k_M} p_{k_M}(k_B-k_C)h'_B(1)+\sum\limits_{k_M}
p_{k_M}k_Ch'_C(1).\end{split}\end{equation} In Eq.(7), functions
$h'_A(1)$, $h'_B(1)$ and $h'_C(1)$ can be derived from Eqs.(3)-(5).
Taking derivatives on both sides of Eqs.(3)-(5) at $x = 1$, we have
\begin{equation}
\label{cases}h'_A(1) =
\lambda_A+\lambda_A<k_M>^{-1}(m_{11}h'_A(1)+m_{12}h'_B(1)+m_{13}h'_C(1)),\end{equation}
\begin{equation}
\label{cases}h'_B(1) =  \lambda_B+
\lambda_B<k_M>^{-1}(m_{21}h'_A(1)+m_{22}h'_B(1)+m_{23}h'_C(1)),\end{equation}
\begin{equation}
\label{cases}h'_C(1) =  \lambda_C+
\lambda_C<k_M>^{-1}(m_{31}h'_A(1)+m_{32}h'_B(1)+m_{33}h'_C(1)),\end{equation}
where
\begin{equation*}<k_M>=\sum\limits_{k_M} |k_M|p_{k_M},\end{equation*}
\begin{equation*}m_{11}=\sum\limits_{k_M,k_A-k_C\geq1}|k_M|p_{k_M}(k_A-k_C-1),\end{equation*}
\begin{equation*}m_{12}=\sum\limits_{k_M,k_A-k_C\geq1} |k_M|p_{k_M}(k_B-k_C),\end{equation*}
\begin{equation*}m_{13}=\sum\limits_{k_M,k_A-k_C\geq1} |k_M|p_{k_M}k_C,\end{equation*}
\begin{equation*}m_{21}=\sum\limits_{k_M,k_B-k_C\geq1} |k_M|p_{k_M}(k_A-k_C),\end{equation*}
\begin{equation*}m_{22}=\sum\limits_{k_M,k_B-k_C\geq1}|k_M|p_{k_M}(k_B-k_C-1),\end{equation*}
\begin{equation*}m_{23}=\sum\limits_{k_M,k_B-k_C\geq1} |k_M|p_{k_M}k_C,\end{equation*}
\begin{equation*}m_{31}=\sum\limits_{k_M,k_C\geq1} |k_M|p_{k_M}(k_A-k_C),\end{equation*}
\begin{equation*}m_{32}=\sum\limits_{k_M,k_C\geq1} |k_M|p_{k_M}(k_B-k_C),\end{equation*}
\begin{equation*}m_{33}=\sum\limits_{k_M,k_C\geq1} |k_M|p_{k_M}(k_C-1).\end{equation*}
From Eqs.(8)-(10), we have
\begin{equation}Mh=-<k_M>e,\end{equation} where
$$M=\left (\begin
{array}{ccc}
-\lambda_A^{-1}<k_M>+m_{11} & m_{12} & m_{13}\\
m_{21} & -\lambda_B^{-1}<k_M>+m_{22} & m_{23}\\
m_{31} & m_{32} & -\lambda_C^{-1}<k_M>+m_{33}\\
\end{array} \right ),$$\\
$h=(h'_A(1)\ h'_B(1)\ h'_C(1))^T,$ and $\ e=(1\ 1\ 1)^T$. Therefore,
$h'_A(1)$, $h'_B(1)$, $h'_C(1)$ diverge at the point where
\begin{equation}\emph{\emph{det}}M= 0,\end{equation} from which we can
calculate the set of the epidemic thresholds
$\{\lambda_{Tc}=(\lambda_{TAc},\lambda_{TBc})\}$ of the TPN.

\subsection{Calculations of outbreak size}
When an epidemic spread across the TPN, the infected nodes will form
into a giant component. Let $u_A$, $u_B$ and $u_C$ be the average
probabilities that a node is not connected to the giant component
via the edge-A, edge-B and edge-C, respectively. According to
percolation theory there are two ways this can happen: either the
edge in question can be unoccupied, or it is occupied but the node
at the other end of the edge is itself not a member of the giant
component. The latter happens only if that node is not connected to
the giant component via any of its other edges. Thus we have
\begin{equation}\begin{split}
\label{cases}u_A=1-\lambda_A+ \lambda_A\times
\frac{\sum\limits_{k_M,\ k_A-k_C\geq1}
|k_M|p_{k_M}u_A^{k_A-k_C-1}u_B^{k_B-k_C}
u_C^{k_C}}{\sum\limits_{k_M} |k_M|p_{k_M}},\ \
\end{split}\end{equation}
\begin{equation}\begin{split}
\label{cases}u_B = 1-\lambda_B+ \lambda_B\times
 \frac{\sum\limits_{k_M,\ k_B-k_C\geq1}
|k_M|p_{k_M}u_A^{k_A-k_C}u_B^{k_B-k_C-1}
u_C^{k_C}}{\sum\limits_{k_M} |k_M|p_{k_M}},\end{split}\end{equation}
\begin{equation}\begin{split}
\label{cases}u_C = 1-\lambda_C+ \lambda_C\times
\frac{\sum\limits_{k_M,\ k_C\geq1}
|k_M|p_{k_M}u_A^{k_A-k_C}u_B^{k_B-k_C}
u_C^{k_C-1}}{\sum\limits_{k_M}
|k_M|p_{k_M}}.\end{split}\end{equation} For the whole TPN, the
outbreak size, i.e. the size of the giant component can be
calculated by
\begin{equation} \label{cases}s = 1-\sum\limits_{k_M}
p_{k_M}u_A^{k_A-k_C}u_B^{k_B-k_C} u_C^{k_C}.\end{equation}
\textbf{Note}: The traditional single-vector propagation model is a
special case of our proposed two-vectors propagation model. The
epidemic threshold $\lambda_A={<k_A>}/({<k^2_A>-<k_A>})$ [7] and the
outbreak size $s = 1-\sum\limits_{k_A} p_{k_A}u_A^{k_A}$ [7] of
single-vector propagation network can be obtained from Eq.12 and
Eq.16 when $k_B=k_C=0$. That is to say, the results of this letter
are applicable in a more general situation.

\section{Simulation results and discussions}
\label{} In this section, we evaluate the two-vectors propagation of
epidemics over the TPN by simulations. Three different types of TPN
are constructed where ($i$) both of these two SPNs are scale-free
(SF) networks; ($ii$) one SPN is Erd\H{o}s-R\'{e}nyi (ER) random
network and the other one is SF network; ($iii$) both of them are ER
networks. We use `X(a,b)' to describe a SPN, where X is the type of
network, `a' is the network size and `b' is the average degree. For
example, SF(2000,3) denotes a SPN comprised of 2000 nodes with
average degree 3. For convenience, we term these three TPNs as
SF-SF, ER-SF and ER-ER, respectively.

\subsection{Epidemic threshold and outbreak size}

In Fig.2, the three-dimensional(3D) curved surfaces obtained by
simulations, indicate the outbreak sizes for epidemic spreading
corresponding to different combinations of spreading rates
$(\lambda_A,\lambda_B)$, where $\lambda_A$ and $\lambda_B$ are the
respective spreading rates of epidemic over two SPNs. The vertical
fences obtained by calculating Eq.(12), represent the theoretical
epidemic thresholds of the epidemic which spreads over TPN. We can
see that the theoretical epidemic thresholds are accurate in judging
the endemic state. It can be found obviously that an epidemic could
spread across the TPN even if these two SPNs well below their
respective epidemic thresholds, which is also independent of the
construction of the TPN. Let $\lambda_{Ac}$ and $\lambda_{Bc}$ be
the respective epidemic thresholds of two SPNs. As shown in
Fig.2(a), the epidemic threshold $\lambda_{Ac}$ of SF(2000,3.997)
corresponds to the epidemic threshold
$(\lambda_{TAc},\lambda_{TBc})$ of SF(2000,3.997)-SF(2000,3.998)
where $\lambda_{TBc}=0$, hence $\lambda_{Ac}=0.1056$. Similarly, we
get $\lambda_{Bc}=0.1102$. From Fig.2 we can see that, for any
epidemic threshold $(\lambda_{TAc},\lambda_{TBc})$ of
SF(2000,3.997)-SF(2000,3.998) where
$\lambda_{TAc}\cdot\lambda_{TBc}\neq 0$, there are
$\lambda_{TAc}<\lambda_{Ac}$ and $\lambda_{TBc}<\lambda_{Bc}$.
\begin{figure}[!htb]
\begin{tabular}{ccc}
\begin{minipage}{0.32\linewidth}
\centering
\includegraphics[totalheight=4.1cm]{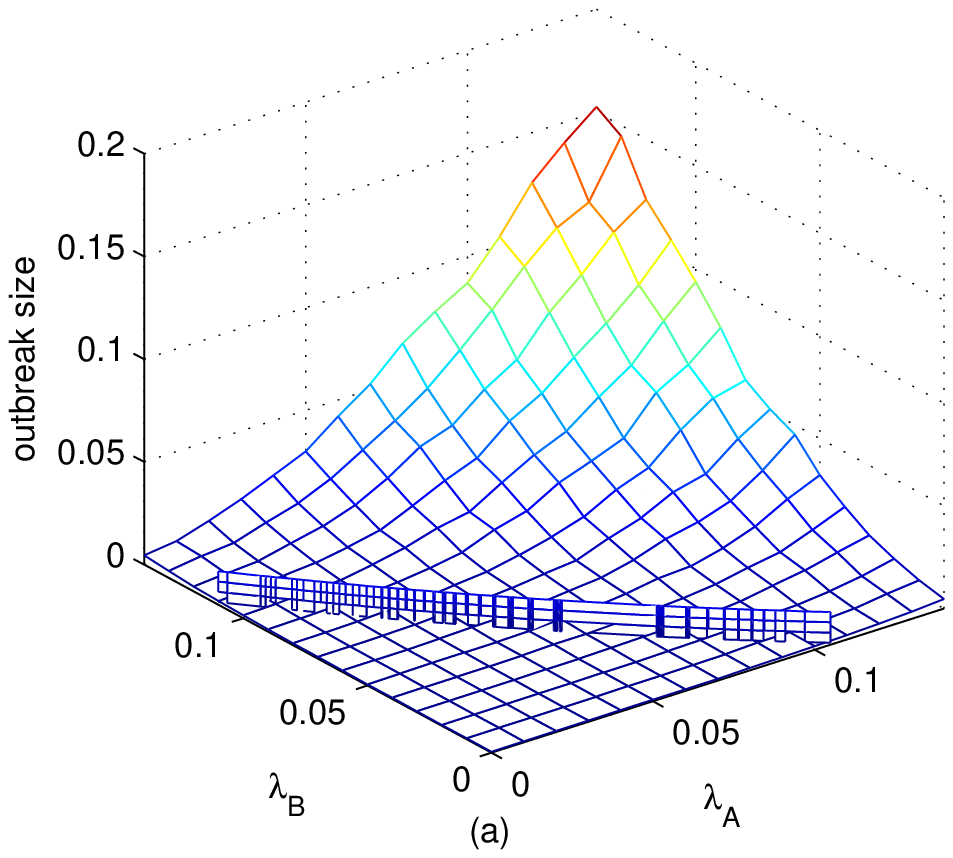}
\end{minipage}
\begin{minipage}{0.32\linewidth}
\centering
\includegraphics[totalheight=4.1cm]{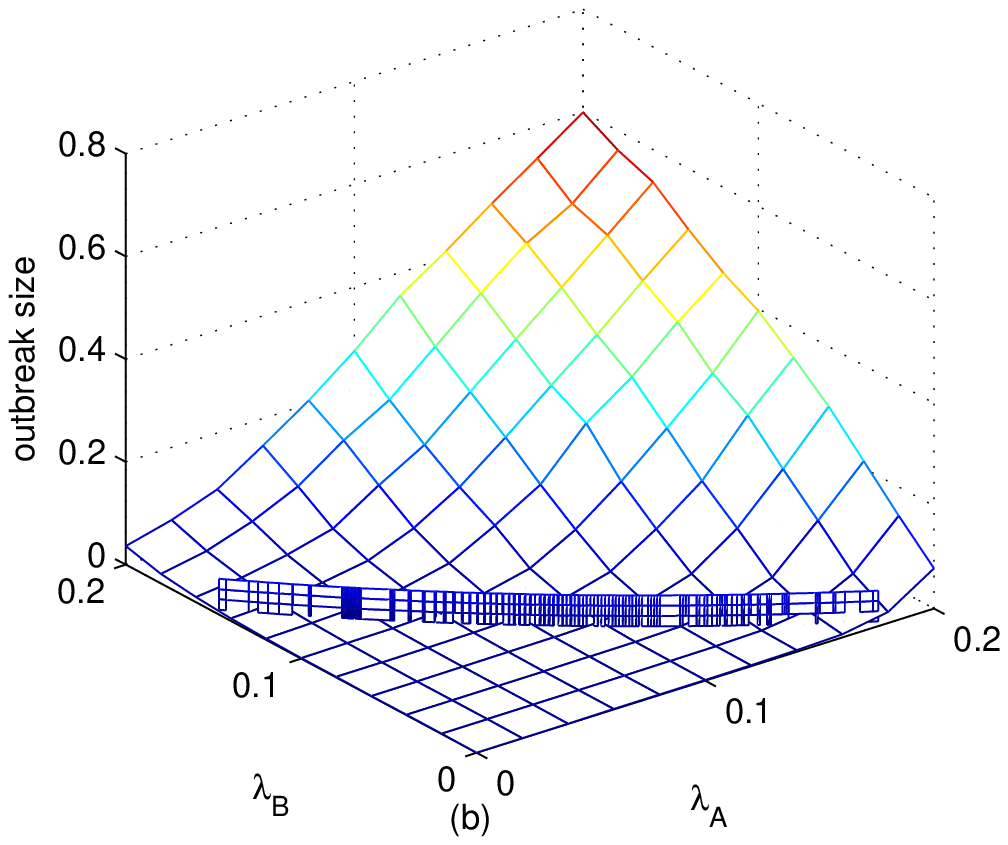}
\end{minipage}
\begin{minipage}{0.32\linewidth}
\centering
\includegraphics[totalheight=4.1cm]{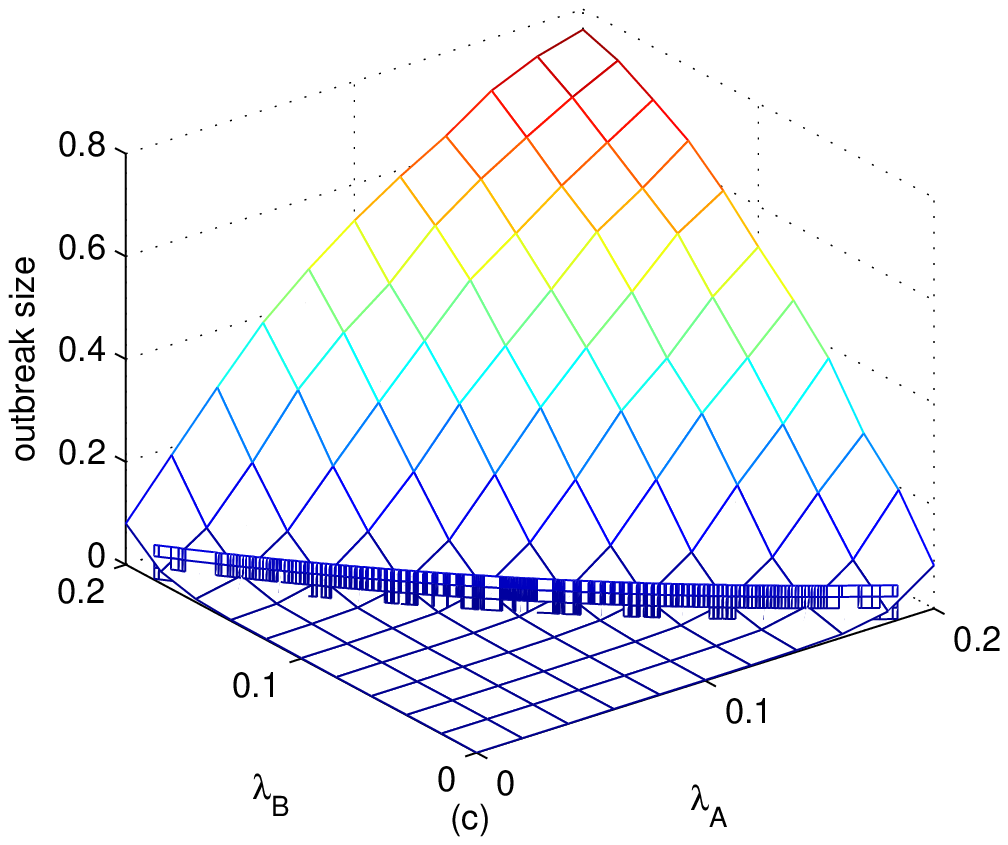}
\end{minipage}
\end{tabular}
\caption{3D curved surfaces obtained by simulations, indicate the
outbreak sizes. The vertical fence represents the theoretical
epidemic threshold. Three TPNs models are (a)
SF(2000,3.997)-SF(2000,3.998); (b) ER(2000,5.883)-SF(2000,3.997) and
(c) ER(2000,5.922)-ER(2000,5.965), respectively.}
\end{figure}

We now evaluate the accuracy of our theoretical derivation for the
outbreak size by comparing the results obtained by theoretical
analysis and simulations. There are two 3D curved surfaces in each
3D map of Figs.3(a-c), where the light one indicates the theoretical
outbreak size and the dark one represents the experimental value. To
get a clearer sight, three section planes of each 3D map are shown
in Figs.3(d-f). We observe that theoretical calculations are in
correspondence with  the data of experiments.
\begin{figure}[!htb]
\begin{tabular}{ccc}
\begin{minipage}{0.32\linewidth}
\centering
\includegraphics[totalheight=8.5cm]{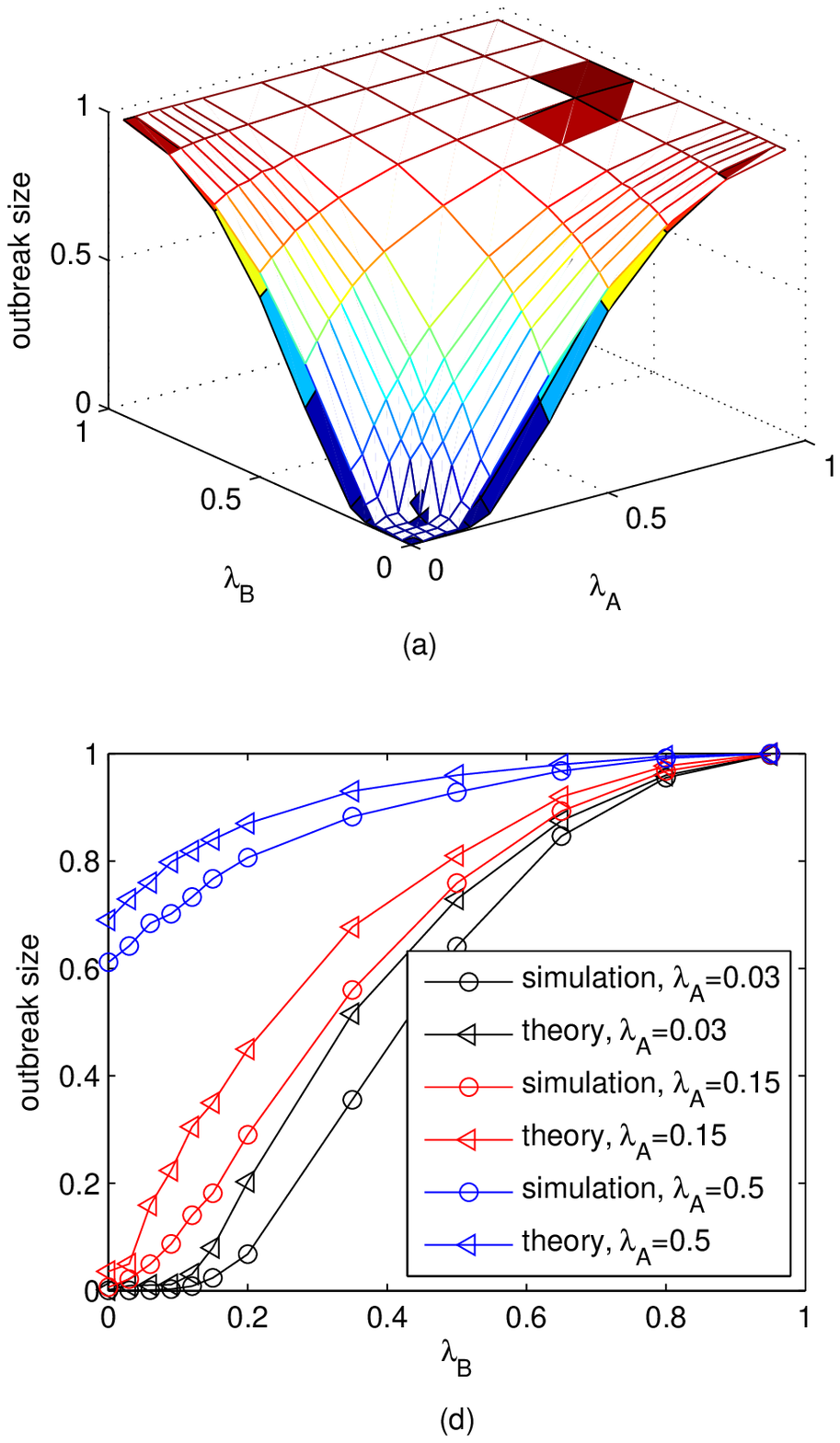}
\end{minipage}
\begin{minipage}{0.32\linewidth}
\centering
\includegraphics[totalheight=8.5cm]{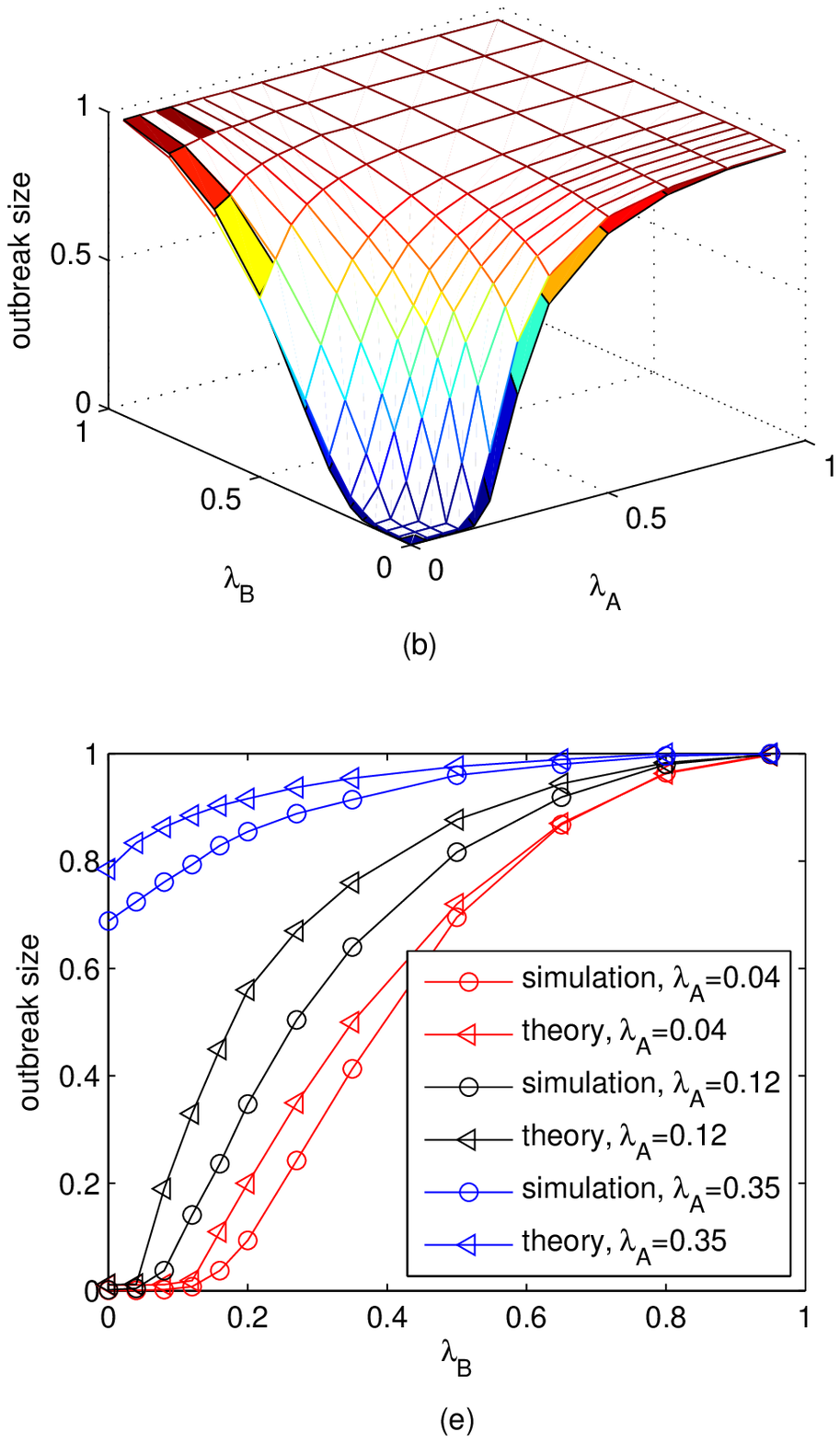}
\end{minipage}
\begin{minipage}{0.32\linewidth}
\centering
\includegraphics[totalheight=8.5cm]{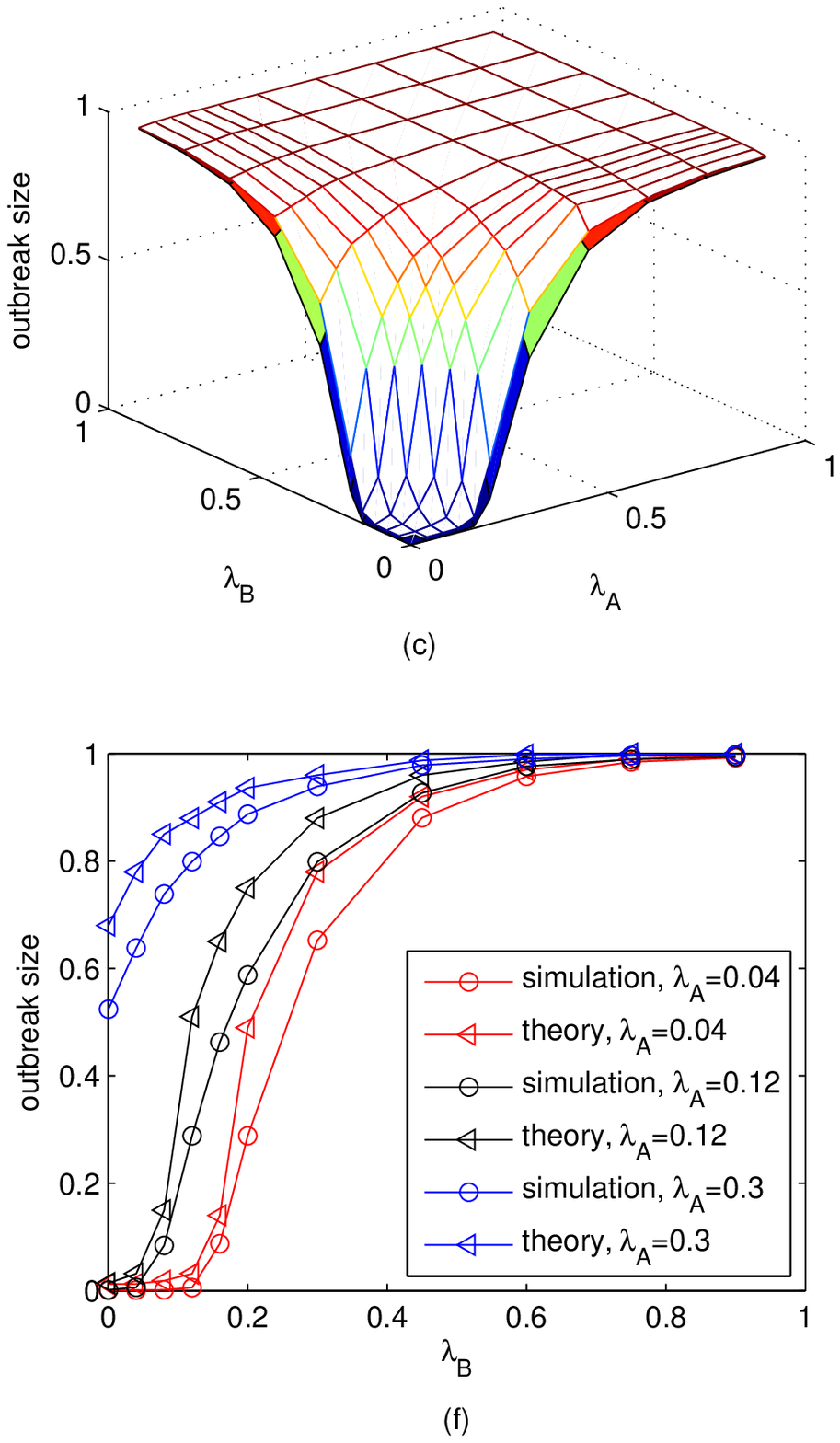}
\end{minipage}
\end{tabular}
\caption{In (a-c), the light and dark 3D curved surfaces are the
theoretical and experimental outbreak sizes, respectively. (d-f)
show some corresponding section planes of (a-c). Three TPNs models
are (a,d) SF(2000,3.997)-SF(2000,3.998); (b,e)
ER(2000,5.883)-SF(2000,3.997) and (c,f)
ER(2000,5.922)-ER(2000,5.965), respectively.}
\end{figure}

\subsection{ASN and DDC}

In the real world, an individual of the TPN may have some same
neighbors in the two SPNs. In section 2, we used a measure ASN to
assess the average similarity between the neighbors from different
SPNs of nodes in the TPN. Another quantity DDC is also developed to
describe the degree-degree correlation of nodes in different SPNs.
Positive values of DDC indicate that high degree nodes in one SPN
are also high degree nodes in the other SPN and vise versa. While
the topologies of the SPNs remain unchanged, the topologies of the
TPN could be affected to some degree by the ASN and DDC, and may
ultimately influence the epidemic processes of the epidemic over
TPN.

It is easy to achieve any targeted value of ASN for SF-SF and ER-ER.
Two SF(ER)-SPNs can be obtained by preferentially(randomly) adding
edges to a same SF(ER) network which has been constructed,
respectively. The value of ASN is determined by the number of the
edges added and the edges of the initial network. It is hard,
however, to achieve a large range of ASN for ER-SF. Assume that the
nodes have same tabs in each SPN, then we can get some different
values of ASN for two SPNs by randomly exchanging the tabs of nodes
for one SPN. In the ER-SF model, we used in Figs.4(b,e), the value
of ASN roughly lies in the interval of [0.02 0.14], while for the
SF-SF and ER-ER models, the corresponding intervals are all [0, 1].

In the following experiments, we assume the epidemic has same
spreading rates when propagates in the two SPNs, that is
$\lambda_A=\lambda_B$. As shown in Fig.4, the epidemic threshold and
the outbreak size are barely affected by the ASN no matter what
types of the two SPNs. This can be understood that when the
topologies of the SPNs remain unchanged, high ASN means nodes can
affect more of their neighbors with the large spreading rate
$\lambda_C$, but the average number of their neighbors is relatively
small. Similarly, although low ASN implies the nodes have more
neighbors, most of the spreading rates between neighbors are the
relatively small $\lambda_A$ and $\lambda_B$. In such cases, the
average number of new infected nodes at a time step may equal no
matter what the values of ASN.
\begin{figure}[!htb]
\begin{tabular}{ccc}
\begin{minipage}{0.32\linewidth}
\centering
\includegraphics[totalheight=8.5cm]{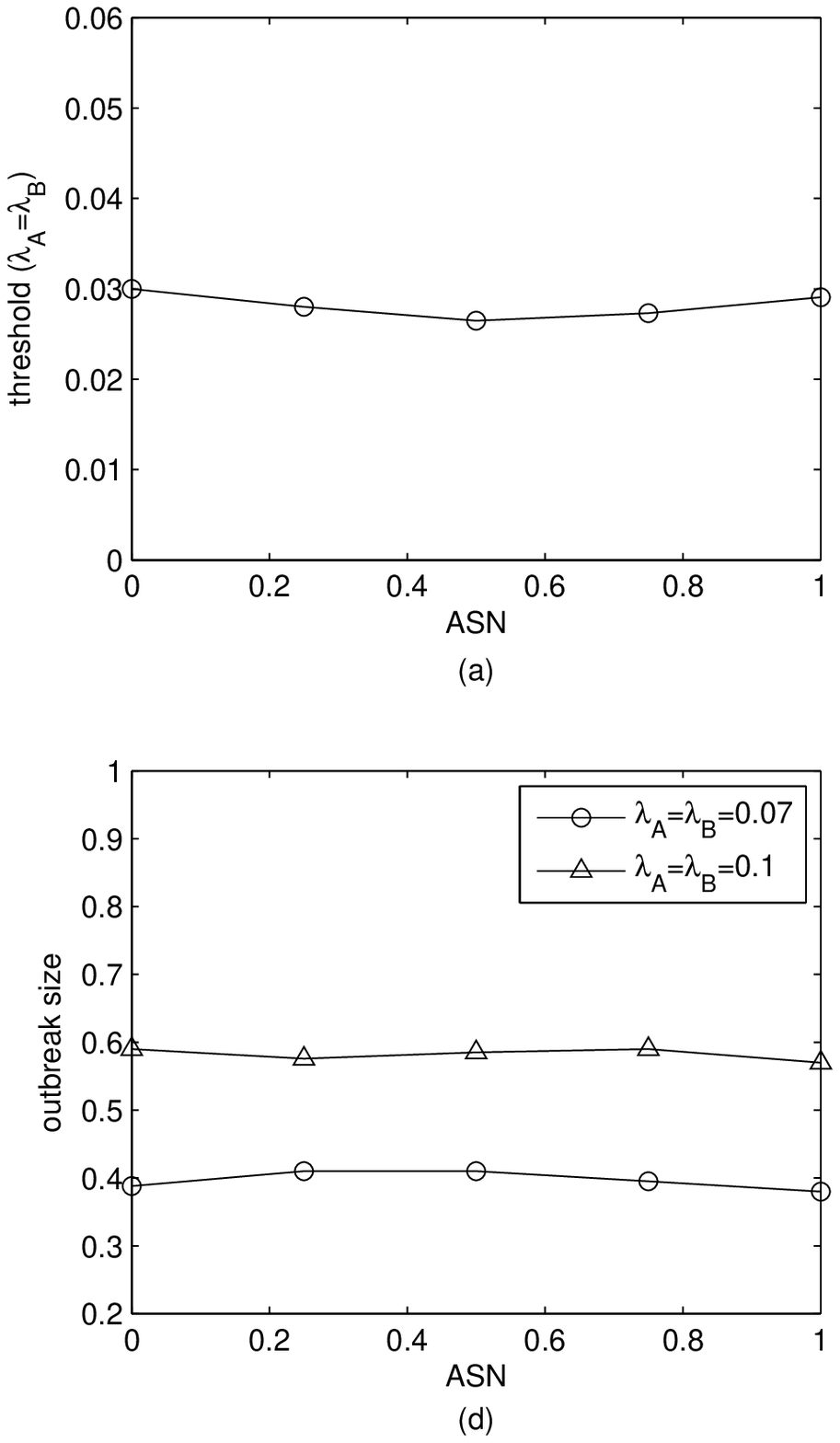}
\end{minipage}
\begin{minipage}{0.32\linewidth}
\centering
\includegraphics[totalheight=8.5cm]{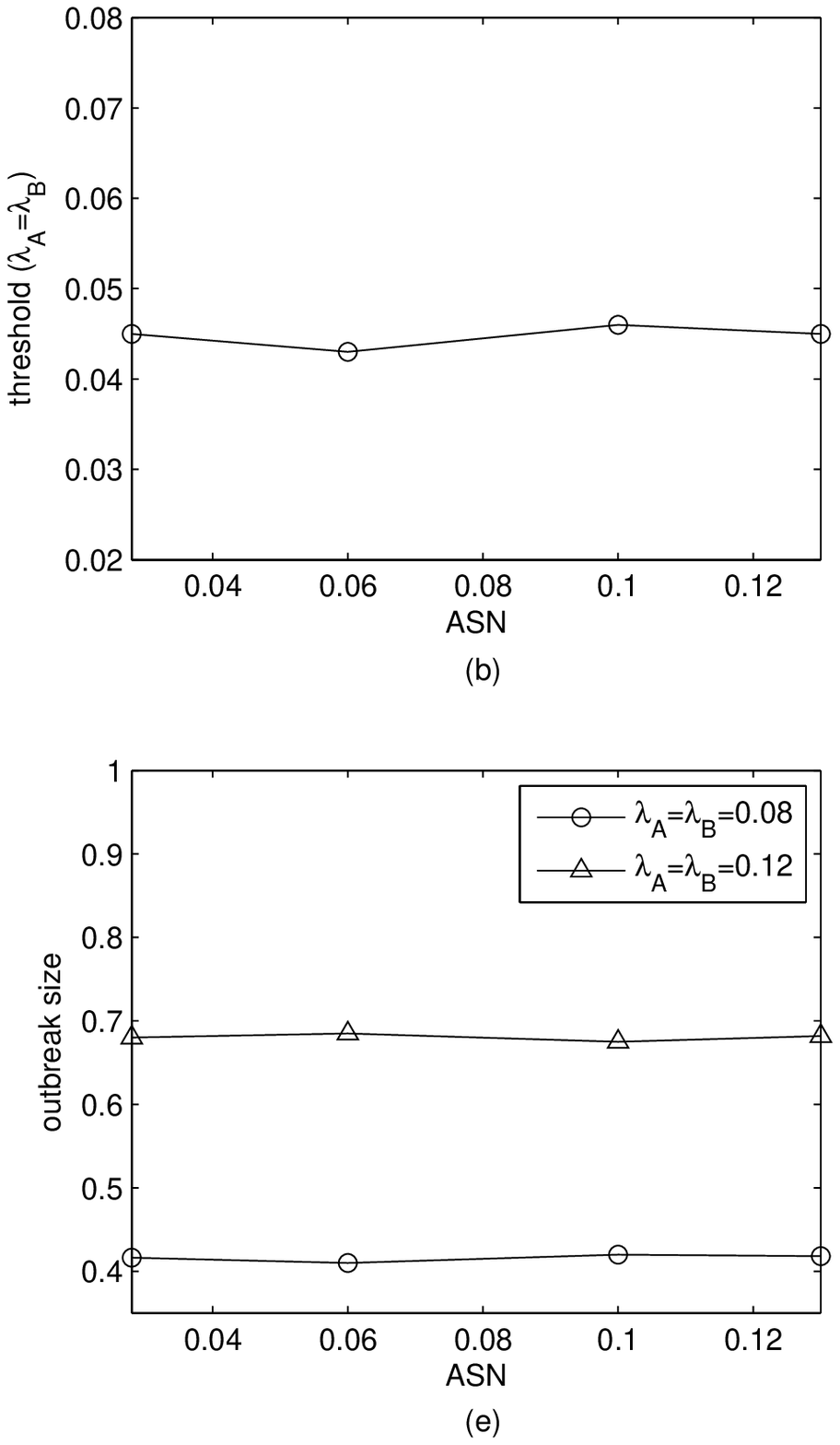}
\end{minipage}
\begin{minipage}{0.32\linewidth}
\centering
\includegraphics[totalheight=8.5cm]{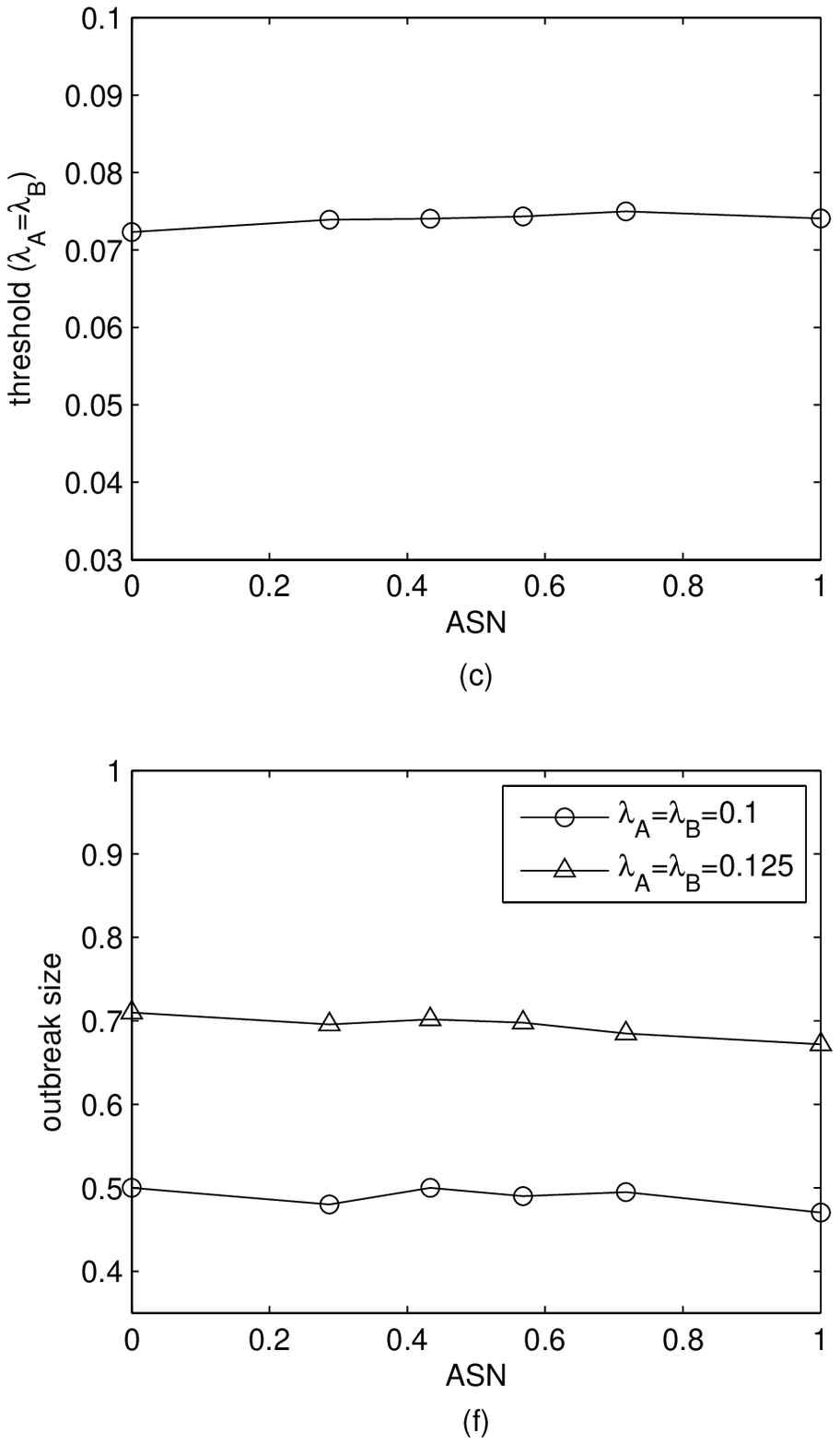}
\end{minipage}
\end{tabular}
\caption{Epidemic threshold and outbreak size for epidemic spreading
in (a,d) SF-SF; (b,e) ER-SF; and (c,f) ER-ER.}
\end{figure}

Fig.5 shows the influences of DDC on the epidemic threshold and the
outbreak size of the epidemics. The different values of the DDC
between two SPNs are achieved by randomly exchanging the tabs of
nodes of one SPN. As shown in Fig.5, a higher DDC can lead to a much
lower epidemic threshold  and a relatively smaller outbreak size no
matter what topologies of the SPNs. The reasons can be explained as
follows: high DDC means high degree nodes in one SPN are also high
nodes in the other SPN and low degree nodes in one SNP also low
degree nodes in the other SPN which leads to increased differences
between the degrees of nodes in TPN. Instead, low DDC leads to
decreased differences between the degrees of nodes in TPN. That is,
high DDC makes the SF-SF and ER-SF the strengthened inhomogeneous
networks and ER-ER a proximate inhomogeneous network, low DDC
however makes the different types of TPNs the proximate homogeneous
network. We have known that [23], the epidemic in the inhomogeneous
network has a faster spread since the existence of high degree nodes
and smaller outbreak size since the low degree nodes are not prone
to be infected, than in the homogeneous network when these two
networks have same average degree. This theory perfectly explains
the results of our experiments.
\begin{figure}[!htb]
\begin{tabular}{ccc}
\begin{minipage}{0.32\linewidth}
\centering
\includegraphics[totalheight=8.5cm]{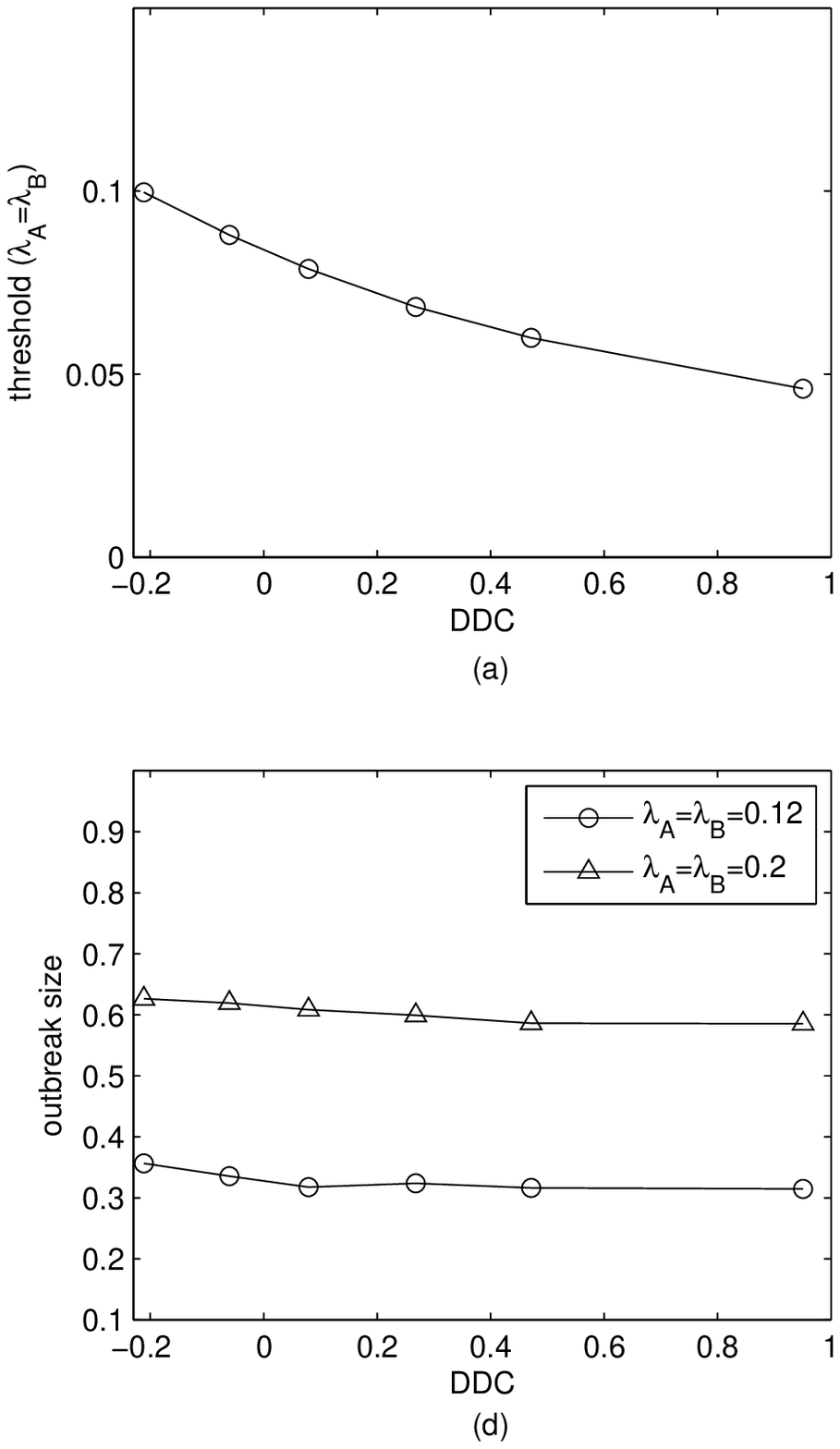}
\end{minipage}
\begin{minipage}{0.32\linewidth}
\centering
\includegraphics[totalheight=8.5cm]{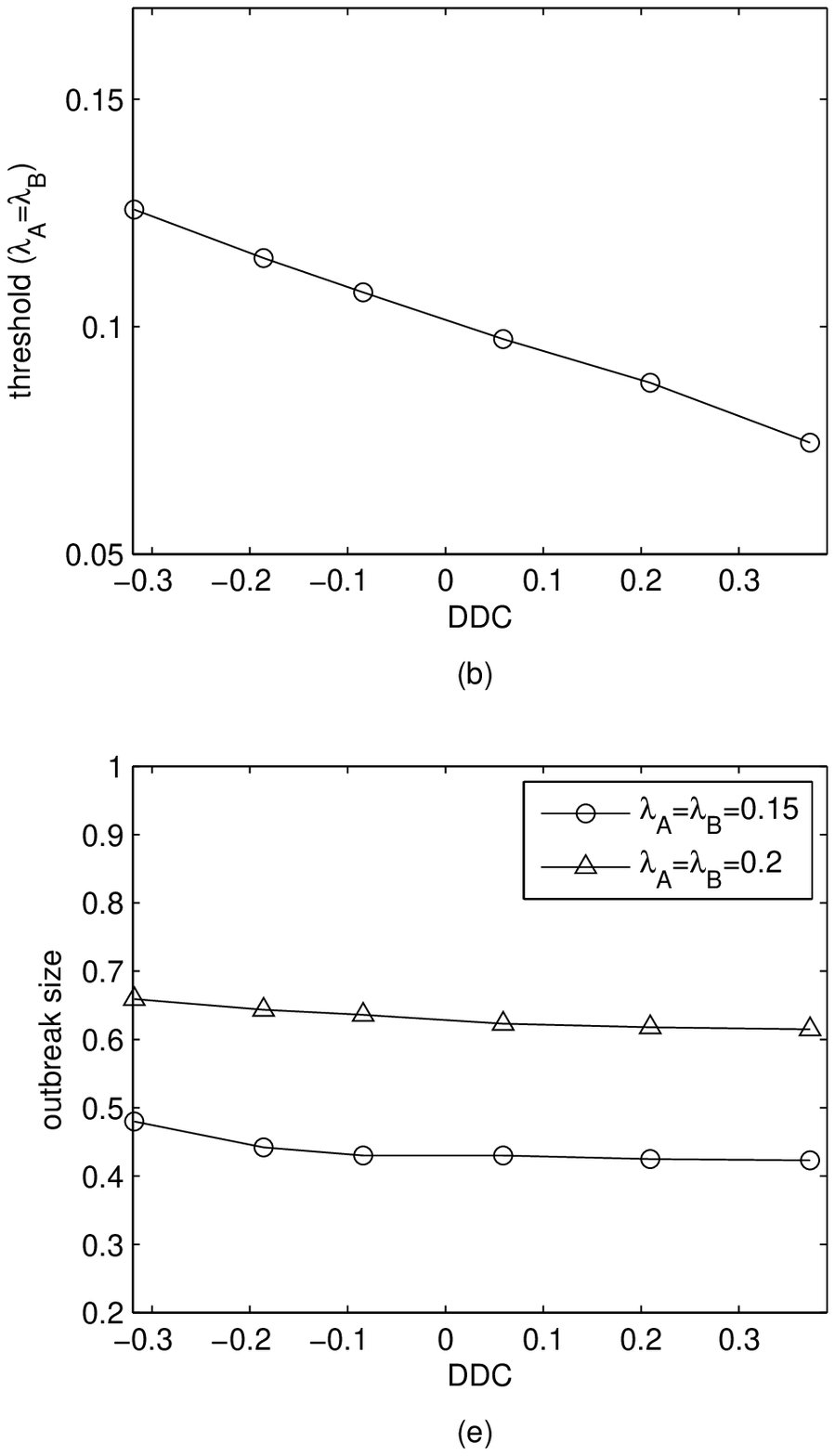}
\end{minipage}
\begin{minipage}{0.32\linewidth}
\centering
\includegraphics[totalheight=8.5cm]{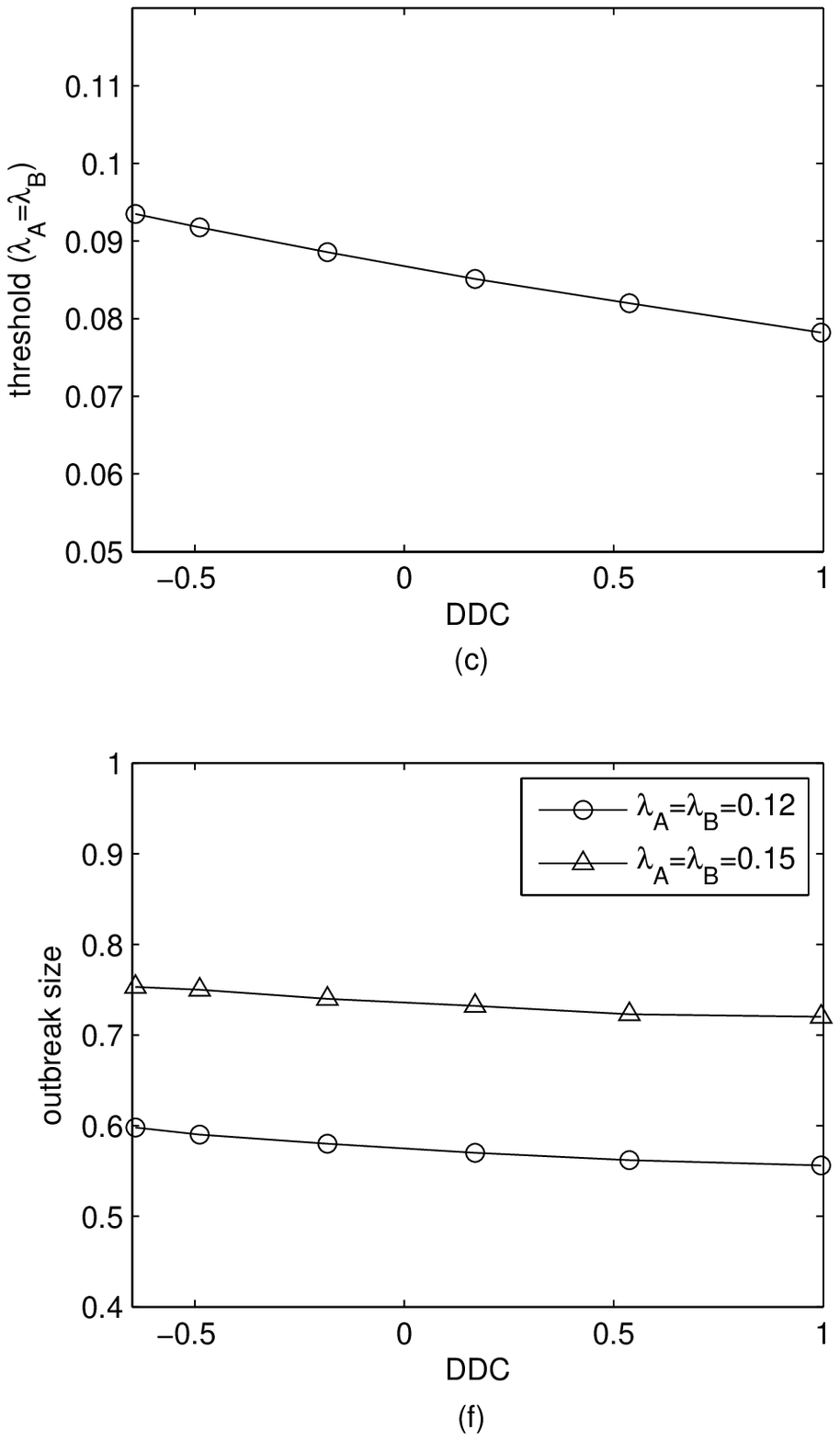}
\end{minipage}
\end{tabular}
\caption{Epidemic threshold and outbreak size for epidemic spreading
in (a,d) SF(2000,3.997)-SF(2000,3.995); (b,e)
ER(2000,4.005)-SF(2000,3.997); and (c,f)
ER(2000,5.950)-ER(2000,5.956).}
\end{figure}

\section{Conclusions}
\label{} In this letter, we demonstrated the dynamics of two-vectors
propagation of epidemics over TPN . Our main contributions can be
summarized as follows: (1) We presented the multiple-vectors
propagation system of epidemics and derived equations to accurately
calculate the epidemic threshold and outbreak size in the TPN. (2)
We found that the epidemics could spread across the TPN even if two
SPNs are well below their respective epidemic thresholds. (3) We
proposed two quantities for measuring the level of inter-similarity
between two SPNs. ASN evaluates the average similarity between the
neighbors from different SPNs of nodes in the TPN which is found
barely affect the epidemic threshold and outbreak size of epidemics
in the TPN. DDC describes the degree-degree correlation of nodes in
different SPNs. It is found that a higher DDC could lead to a much
lower epidemic threshold  and a relatively smaller outbreak size no
matter what topologies of the SPNs.

Although we consider the epidemics spreading on TPN which superposed
by only two SPNs, it is easily to extend the model to an arbitrary
number of SPNs with any size. Our research not only provide useful
tools and insights for further studies of dynamics of multi-vectors
propagation of epidemics, but also has important implications for
the design of efficient control strategies.

\section{Acknowledgement}
This paper was supported by the Foundation for the Author of
National Excellent Doctoral Dissertation of PR China (Grant No.
200951), the National Natural Science Foundation of China (Grant
Nos. 61100204, 61170269, 61121061), the Asia Foresight Program under
NSFC Grant (Grant No. 61161140320).

\end{document}